\documentclass{ws-procs9x6}

\begin{document}

\title{PROGRESS AND ISSUES IN THE ELECTROMAGNETIC 
  PRODUCTION OF KAON ON THE NUCLEON}

\author{T. MART}

\address{Departemen Fisika, FMIPA, Universitas Indonesia
Depok 16426, Indonesia\\
tmart@fisika.ui.ac.id}

\begin{abstract}
The present status of kaon photo- and electroproduction on 
the nucleon is briefly reviewed. Some current important issues in 
this field are discussed.
\end{abstract}

\keywords{Kaon, photoproduction, electroproduction, nucleon resonance}

\bodymatter

\section{Introduction}\label{sec:intro}
Compared to the case of pions the electromagnetic productions of
kaons are less understood. This is due to the fact that 
the strangeness quantum
number is explicitly present in the final state of the process.
Although the reactions involving kaons are slightly more 
complicated, the additional 
degrees of freedom created by the strangeness can give information 
not available from nucleons and pions. For instance, due to the
conservation of the strangeness, the production of kaons is 
always accompanied by the creation of hyperons,
which can be used to deeply explore the structure of the nucleus 
since they are not blocked by the Pauli principle. On the other hand,
this elementary process provides an important input for 
the calculation of hypernuleus photo- or electroproduction. 
Along with the fact that some $N^*$
resonances that were predicted in quark models have only 
noticeable branching ratios into the $K \Lambda$ channel,  
kaon photo- and electroproduction 
clearly have drawn many attentions for more than five decades.

\section{Theoretical Models}
The earliest attempt to theoretically explain kaon 
photo- and electroproduction was proposed by 
Kawaguchi and Moravcsik more than 
50 years ago\,\cite{masasaki}. Interestingly, 
all of the six possible isospin channels were
already considered by utilizing only three Feynman diagrams 
of the Born terms, though hitherto no experimental data 
were available. The results were obviously very modest and, 
in fact, the cross sections for the $K^0\Lambda$ and $K^0\Sigma^0$ 
channels were predicted to be zero since neither nucleon 
anomalous magnetic moments nor resonance contributions 
were taken into account. Since that time considerable 
efforts were devoted to explain the appearing experimental 
data, although only Thom who tried to seriously fit the data
to an isobar model. After that, there were a number of works
using isobar model\,\cite{deo}, dispersion 
relation\,\cite{nelipa,pickering}, multipoles 
analysis\,\cite{schorsch}, and Regge approach\,\cite{levyph}. 
After the mid-seventies, the interest in this field  
was temporarily dormant, 
mainly due to the lack of experimental facilities. 

The interest in kaon photo- and electroproduction 
was revived by the constructions of 
modern accelerators such as those in JLAB, MAMI, 
ELSA, and others laboratories. The work of 
Adelseck, Bennhold, and Wright started the new
era of phenomenological models in $K^+\Lambda$ 
photoproduction\,\cite{abw}. This work was 
refined with the inclusion of more data and 
electroproduction process\,\cite{adel2} and
extended to all isospin channels\,\cite{terry1}.
A chiral quark model with less parameters has been
also put forward\,\cite{zpli} in an attempt to
recover the low energy theorem. The results 
of this model are modest and, in fact, the $K^0\Sigma^+$
cross section is predicted to be larger than that for  
$K^+\Sigma^0$ by a factor of approximately two, since 
contributions  from the seagull and $s$-channel 
resonance have opposite signs. 

Modern calculations of kaon photoproduction exploit
chiral perturbation theory\,\cite{chpt}
and coupled channels analysis\,\cite{coupled-channels,coupled}. 
In the higher energy regions, the Regge\,\cite{Guidal:1997hy} or 
hybrid models\,\cite{hybrid} (a combination of the Feynman 
diagrammatic technique and Regge formalism) turn out
to be more appropriate. 
Nevertheless, for practical use such as
for nuclear applications and other phenomenological studies,
the single channel isobaric analysis is still proven to
be powerful\,\cite{terry1,Mart:1999ed,single_channel}.

\section{Experimental Data}
\label{sec:data}
\begin{figure}[t]
\psfig{file=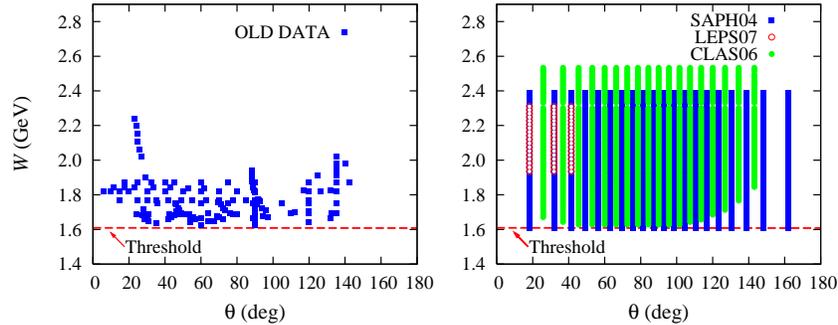,width=4.5in}
\caption{Kinematical coverages of the experimental data from
  old databases (left) and new experiments (right) for the
  $\gamma p \to K^+ \Lambda$ differential cross section.
  In the right panel data from SAPHIR\,\cite{glander04} (SAPH04),
  CLAS\,\cite{bradford06} (CLAS06), and 
  LEPS\,\cite{sumihama06} (LEPS07) collaborations are shown.}
\label{fig:data}
\end{figure}

Historically, experimental data of kaon photo- and electroproduction
can be divided into two categories, old data (published before
1980) and new data (published after 1990). A list of references
for old data is given, e.g., in Refs.\,\cite{AS90,saphir98},
whereas the corresponding 
kinematical coverage for the photoproduction is shown in
the left panel of Fig.~\ref{fig:data}. Surprisingly, 
the old data cover also forward angle regions, the case which
is very difficult to achieve with the presently available 
technologies, as obviously shown in the right panel of Fig.~\ref{fig:data}. 
As discussed in the following section, this
region is in fact very decisive for constraining the models and for 
nuclear applications. Nevertheless, the data quality is
in general poor and, as a consequence, the cross sections data
do not show resonance structures. There were few data available
for the recoiled $\Lambda$ polarization and two data points
on the target asymmetry.

Kaon photoproduction started to become more interesting after
the publication of SAPHIR data\,\cite{saphir98} on the $K^+\Lambda$
and $K^+\Sigma^0$ channels in 1998, since the corresponding error bars 
allow for an identification of resonance structures, especially
in the $K^+\Lambda$ channels. Further analysis of SAPHIR 
data\,\cite{glander04} as well as the new CLAS 
data\,\cite{bradford06} confirmed these structures. Note
that in both SAPHIR and CLAS data sets, there are plenty
of data on the recoil polarization. Together with the new
data from the LEPS\,\cite{sumihama06,hicks07} and 
GRAAL\,\cite{lleres07} collaborations, these data 
provide a strong constraint on the proliferation 
of phenomenological models. In the $K\Sigma$ channels,
new data have also appeared from SAPHIR 
collaboration\,\cite{lawall05} ($K^0\Sigma^+$), 
and from the LEPS collaboration\,\cite{kohri06}
($K^+\Sigma^-$). The ability to reverse the direction of
photon helicity in the SPRING8 has led to the measurements
of the photon asymmetry $\Sigma$ in the $K^+\Lambda$,
$K^+\Sigma^0$, as well as $K^+\Sigma^-$ channels\,\cite{sumihama06,kohri06}.

Since the polarization of the recoiled $\Lambda$ can be obtained
without any additional apparatus, measurements of the double
polarization observables $C_x$ and $C_z$ (or $O_x$ and $O_z$) are possible,
provided that the polarized photon beams are available. 
The first measurement of these observables was performed
by the CLAS collaboration\,\cite{Bradford:2006ba} and 
because the available beams at JLAB 
are circularly polarized, the corresponding observables
are  $C_x$ and $C_z$. Very recently, by utilizing the linearly polarized
photon the GRAAL collaboration has been able to measure the
$O_x$ and $O_z$ observables\,\cite{Lleres:2008em} from
threshold up to $E_\gamma=1.5$ GeV.

The electroproduction experiments  with high statistics
have been also performed
at JLAB\,\cite{mohring03,carman03,ambrozewicz}.  
It is reported that the longitudinal 
and transverse components of the cross section can be
nicely separated\,\cite{mohring03,ambrozewicz}, 
whereas the transferred polarization in 
the $\vec{e}p\to e'K^+\vec{\Lambda}$ process
has been measured\,\cite{carman03}.

\section{Current Important Issues in Kaon Photoproduction}
\subsection{The problem of data discrepancy}
\begin{figure}[t]
\centering\psfig{file=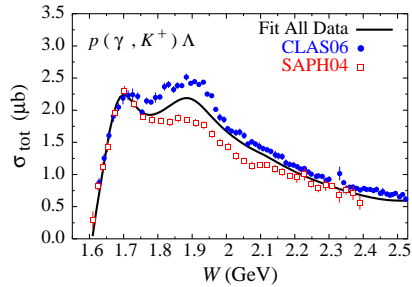,width=2.2in}
\caption{The apparent difference between the 
  CLAS\,\protect\cite{bradford06} and 
  SAPHIR\,\protect\cite{glander04} data
  compared with the result of fitting a multipoles 
  model\,\protect\cite{mart06} to
  both data sets.}
\label{fig:tot}
\end{figure}

In spite of their unprecedented high statistics, the new 
CLAS\,\cite{bradford06} and SAPHIR\,\cite{glander04} data
reveal a lack of mutual consistency in the forward and 
backward regions. This problem can be summarized 
by the total cross section shown in 
Fig.~\ref{fig:tot}, where it is obvious that fitting 
a phenomenological model to both data sets simultaneously results in
a model which is inconsistent to all data sets. As 
shown by Ref.\,\cite{mart06} this problem also  
hinders a reliable extraction of the resonance 
parameters, since the use of SAPHIR and CLAS data, 
individually or simultaneously,
leads to quite different resonance parameters. 
Therefore, the presently available data do  not
allow for a firm conclusion on the extracted 
``missing resonances''.

By using four different isobar models Ref.\,\cite{Bydzovsky:2006wy} 
studied the statistical properties of both CLAS and SAPHIR data.
It is found that the SAPHIR data are coherently shifted 
down with respect to the CLAS and LEPS\,\cite{sumihama06} data, 
especially at forward kaon angles. A global scaling factor of 
15\% is required to remove this discrepancy. The phenomenon also 
implies that the LEPS data are more consistent with the CLAS
data than the SAPHIR ones. Interestingly, at forward angles 
both CLAS and LEPS data sets can be described reasonably well 
within the isobaric model without hadronic form factors.
This can be understood from the energy distribution of
the cross sections at this kinematics, as will be discussed in the
following subsections. The statistical study recommend that
a combination of CLAS, LEPS, and old\footnote{see Section~\ref{sec:data}
for the definition of old data}
data sets in the fitting data base 
is the more preferred choice, while a combination of CLAS and
SAPHIR data should not be assumed in fixing 
parameters of models, especially at forward angles.

\begin{figure}[t]
\centering\psfig{file=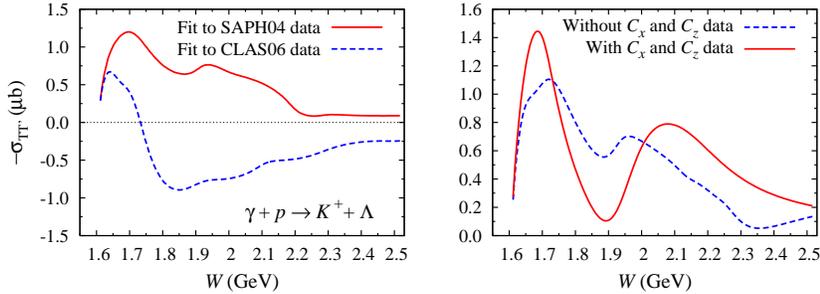,width=4.5in}
\caption{Variations of the total cross section $\sigma_{\rm TT'}$
  obtained from fitting to different experimental data sets (left)
  and fitting to the data sets that include and exclude the
  beam-recoil polarization $C_x$ and $C_z$ data (see 
  Ref.\,\protect\cite{Mart:2008ik} for details).}
\label{fig:sigTT}
\end{figure}

Despite considerable efforts to alleviate this problem, it is
still difficult to determine which data set should be used to 
obtain a reliable phenomenological model. 
The reason is that in all analyses the experimental data are
fitted by adjusting a set of free parameters, while 
the precise values of these parameters are not well known.
Furthermore, the extracted parameters are not unique and
also sensitive to the number of resonances used in a model.
In view of this, Ref.\,\cite{Mart:2008ik} proposed to use
other quantities which
can be predicted by the models and can be directly
compared with the results from other measurements or 
model predictions. One of the possible quantities
is the contribution of the $\gamma p\to K^+\Lambda$ channel
to the Gerasimov-Drell-Hearn (GDH) sum rule  \cite{gdh}, which 
relates the anomalous magnetic 
moment of the proton $\kappa_p$ to the total 
photoabsorption cross sections $\sigma_{\rm TT'}$, i.e.
$\kappa_p^2 = (m_p^2/\pi^2\alpha)
\int_{0}^{\infty}dE_\gamma\,\sigma_{\rm TT'}/ E_\gamma$.
By looking at the energy distributions of the  $\sigma_{\rm TT'}$
obtained from models that fit to CLAS and SAPHIR data shown in
the left panel of Fig.~\ref{fig:sigTT}, it is obvious that the
two data sets lead to different contribution to $\kappa_p$.
It is then concluded that contribution from the model that
fits to the SAPHIR data sets is more preferred\,\cite{Mart:2008ik}.
Nevertheless, experimental measurement of the $\sigma_{\rm TT'}$ is 
clearly required to confirm this claim. Since the SAPHIR detector
has been dismantled, measurements of the $\sigma_{\rm TT'}$ by 
the CLAS or MAMI collaboration seem to be the only choice.

\subsection{The effects of $C_x$ and $C_z$ data}
The beam-recoil polarization observables data, $C_x$ and 
$C_z$, published by  the CLAS collaboration\,\cite{Bradford:2006ba}
recently, indicate that the $\Lambda$ polarization is 
predominantly in the direction of the spin of the 
incoming photon. It is interesting that recent analyses found that 
these data seems to be difficult to explain. 
In Ref.~\cite{Mart:2008ik} it is reported that 
the inclusion of these data reduces the complicated structure
of the total cross sections $\sigma_{\rm TT'}$ as shown in the
right panel of Fig.~\ref{fig:sigTT}, which indicates that 
the $C_x$ and $C_z$ data select only certain important 
resonances, i.e. the $S_{11}(1650)$, $P_{11}(1710)$, 
$P_{13}(1720)$, and $P_{13}(1900)$ resonances. This finding 
corroborates the result of a recent coupled channels 
analysis.~\cite{Anisovich:2007bq}

\subsection{The second peak in the cross section}
The first version of SAPHIR data on $\gamma p\to K^+ \Lambda$ 
released in 1994 did not
give any hint about resonance structures in the cross section,
since the statistics was very limited\,\cite{saphir_old1}.
However, by collecting more statistics the second version
of SAPHIR data\,\cite{saphir98} released in 1998 display 
two distinct peaks in the total cross section 
at $W\approx 1700$ MeV and 1900 MeV. More accurate data 
published by the SAPHIR\,\cite{glander04} and 
CLAS\,\cite{bradford06} collaborations 5 and 8 
years later, respectively, confirm the existence of the two peaks 
(see Fig.\,\ref{fig:tot}). 
By comparing the extracted resonance widths 
and branching fractions with the predictions of a constituent
quark model\,\cite{cqm}, Ref.\,\cite{Mart:1999ed} concluded
that the second peak corresponds to the $D_{13}(1895)$ nucleon
resonance that is missing from the PDG list\,\cite{pdg98}.
Although many subsequent analyses used this spin 3/2 nucleon
resonance in their models, there was a 
suggestion\,\cite{Janssen:2001pe} that this peak could 
come from other resonances such as $S_{11}$, $P_{11}$
or $P_{13}$. However, the possibility of using these
resonances, instead of the  $D_{13}(1895)$, has been also discussed
in Ref.\,\cite{Mart:1999ed}, but it was ruled out after comparing
the extracted widths and branching fractions with the predictions
of quark model. Further study using a multipoles 
approach\,\cite{mart06}
found that, in spite of their differences, both 
SAPHIR\,\cite{glander04} and CLAS\,\cite{bradford06} 
data sets indicate that the peak originates from the 
$D_{13}$ resonance with a mass between 
$1911-1936$ MeV. The necessity of the $D_{13}$ in the analyses
of $K^+\Lambda$ photoproduction with a mass 
of around 1900 MeV has been also corroborated by recent coupled
channels analyses\,\cite{coupled-channels}. Note that in these
analyses the recoil polarization data have been also taken into
account. It is quite interesting, however, that a recent 
effective field theory analysis suggests that the peak
should correspond to the $P_{11}$ state which is a bound
state of $K\bar KN$ with a mixture of $a_0(980)N$ and $f_0(980)N$
components\,\cite{MartinezTorres:2009cw}. To confirm this idea, 
two experiments are proposed, i.e. first to exclude contributions
of the spin 3/2 states and second to find an indication of the
$K\bar KN$ bound state. On the other hand, a preliminary 
study\,\cite{nelson09} 
performed by constraining the free parameters in a multipole 
analysis\,\cite{mart06} to the PDG values indicates that the
peak originates from the contributions of both $S_{11}$ and
$P_{11}$ states with masses around 1920 MeV.
In this study it is found that the contribution of the  $S_{11}$
state is stronger than that of the $P_{11}$ and the result 
does not depend on which data set is being used 
in the fitting database. 

\subsection{The problem at forward angles}
\begin{figure}[t]
\centering\psfig{file=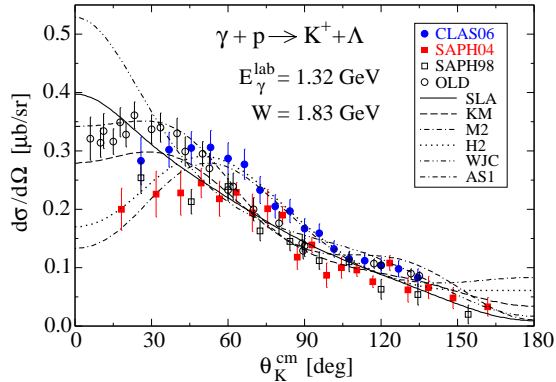,width=2.1in,angle=-90}
\caption{Variations of different models and experimental data
  at forward angles\,\protect\cite{Bydzovsky:2006wy}.}
\label{fig:forward}
\end{figure}

Since the dominant contributions to the total cross section
shown in Fig.~\ref{fig:tot} come from the forward angles data,
the largest discrepancies between the CLAS and SAPHIR data
obviously show up at the forward angles. Unfortunately, 
the available phenomenological models vary wildly at these
kinematical regions. This situation is clearly shown in
Fig.~\ref{fig:forward}, in which we can see that for
$0^\circ\le\theta_K\le 30^\circ$ our best knowledge on
kaon photoproduction cannot tell us about the actual values
of differential cross sections.

Meanwhile, theoretical predictions of the hypernuclear
photo- or electroproduction rely heavily on the elementary
operator extracted from kaon photo- or electroproduction
data at very forward angles. As an example, in the 
electromagnetic production of the hypertriton,
the magnitude of differential cross section is only realistic
for experimental measurement at $0^\circ\le\theta_K\le 20^\circ$,
whereas at $\theta_K\approx 25^\circ$ the cross section is
practically zero\,\cite{Mart:2008gq}. 
In view of this, accurate and reliable experimental
data as well as theoretical formulation of kaon photoproduction
in the forward region 
are the first condition toward a reliable prediction 
of the hypernuclear photoproduction cross section.

\subsection{The controversy of hadronic form factors}
\begin{figure}[t]
\begin{minipage}[!t]{41mm}
\epsfig{file=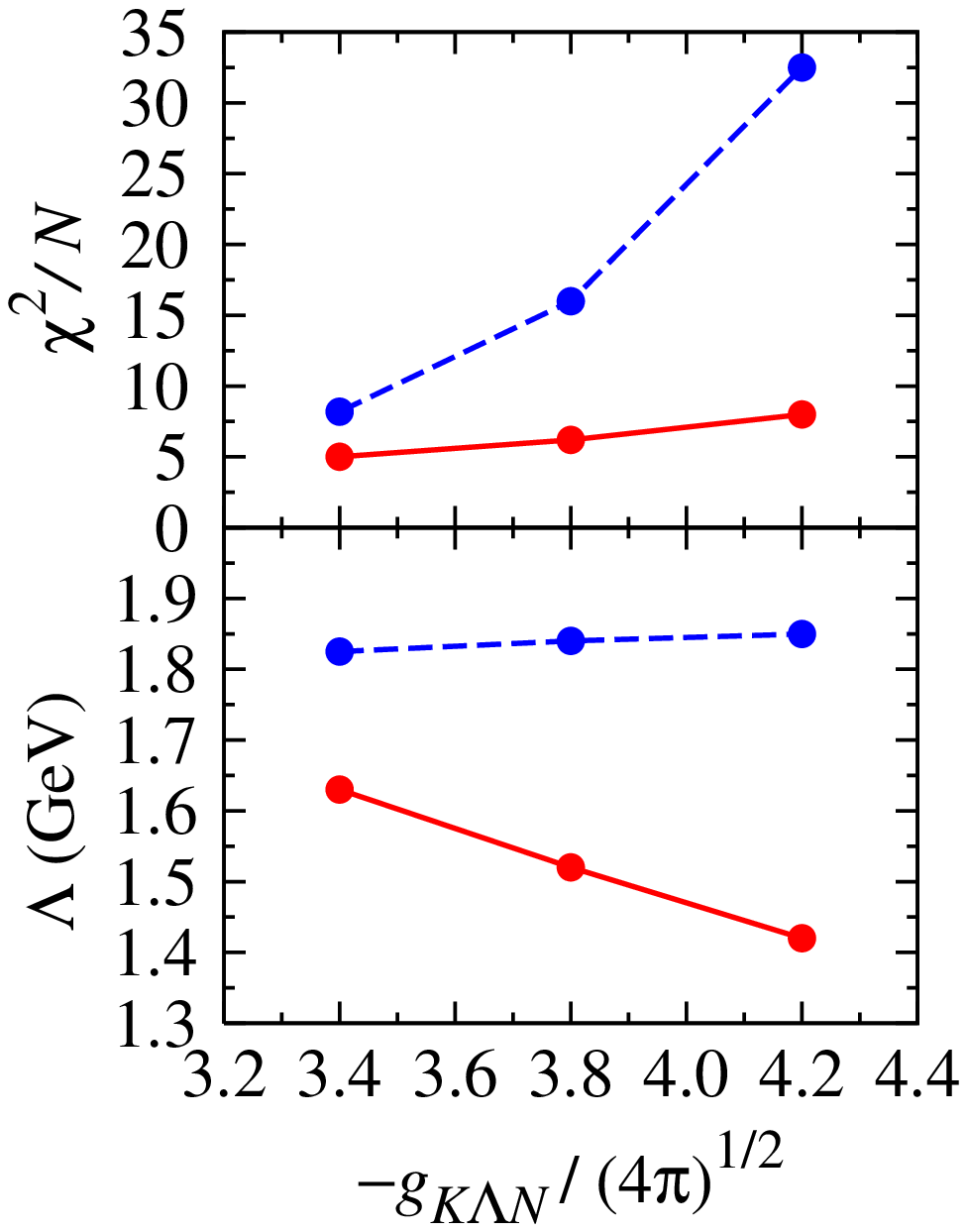, width=41mm}
\end{minipage}
\hspace{1mm}
\begin{minipage}[!t]{48mm}
\epsfig{file=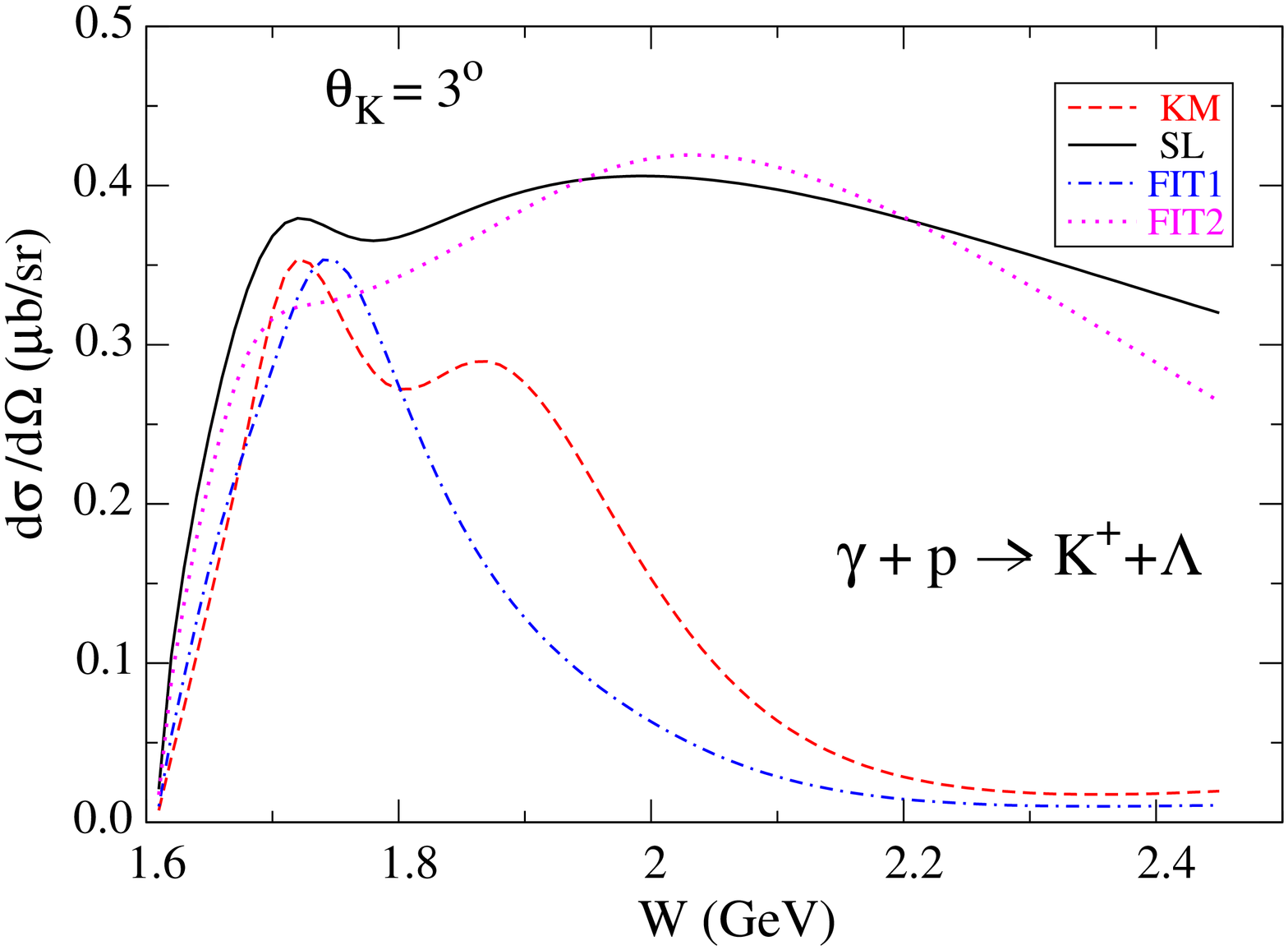, width=48mm, angle=-90}
\end{minipage}
\caption{(Left panel) Variations of the $\chi^2/N$ and the cutoff parameter 
of hadronic form factor $\Lambda$ due to the variation of
the leading coupling constant $g_{K\Lambda N}$ \cite{hbmf98}. 
Solid lines show results obtained with the Haberzettl's method \cite{hh97g} 
whereas dashed lines display the Ohta's prescription \cite{ohta89}.
(Right panel) Effects of the hadronic form factor on the energy 
distribution of the differential cross section \cite{Bydzovsky:2006wy}
for different phenomenological models.
Models of KM and FIT1 utilize the hadronic form factor, while SL and FIT2
exclude this form factor (see Ref.~\cite{Bydzovsky:2006wy} for further
explanation of the used models).\label{fig:had_ff}}
\end{figure}

It has been well known that the background terms 
become divergent at high energies. Therefore,
the use of hadronic form factors to reduce the background
contribution is desired. It is also known that the use of
this form factor leads to the violation of gauge invariance
of the amplitude. As a result, several methods have been
put forward in the last decades to overcome this problem. 
Two of the well known prescriptions are due to 
Ohta \cite{ohta89} and Haberzettl \cite{hh97g}. As can be
seen from the left panel of 
Fig.~\ref{fig:had_ff}, the latter is superior
since it can provide a reasonable description of experimental
data, i.e. producing relatively much smaller $\chi^2$, 
using values for the leading couplings constants 
close to the SU(3) values. In Ohta's method this is not 
possible due to the absence of a hadronic form factor in 
the electric current contribution.

Since then, most of the phenomenological models include hadronic
form factors in their background terms. On the other hand, 
Ref.\,\cite{Janssen:2001pe} proposed to use certain hyperon
resonances, the $S_{01}(1800)$ and $P_{01}(1810)$, for counterbalancing 
the strength of the Born terms. This is achieved through a destructive
interference between these $u$-channel resonances and the Born terms,
thereby reducing the strength to a realistic level. However, the method
is not so convincing for the huge number of presently available 
experimental data and, on the other hand, the use of hadronic form 
factor is more flexible and simultaneously takes into account 
the fact that nucleons are composite objects, and not point-like. 

Recently, Ref.\,\cite{Bydzovsky:2006wy} has demonstrated that
the behavior of forward-angle cross sections as a function of the 
energy for models that include the hadronic form factors is far
from realistic. This finding is depicted in the right panel of
Fig.~\ref{fig:had_ff}, where the hadronic form factors drastically
suppress the cross sections at $W\ge 2$ GeV. 
Although no data are available 
for comparison at this kinematics, such a strong damping has not 
been observed in the experimental data near the forward angles.
As can be seen in Figs.~4 and 6 of Ref.\,\cite{mart06}
differential cross sections for $\theta_K=18^\circ-37^\circ$ 
tend to be flat up to $W=2.5$ GeV. Therefore, the concept 
of hadronic form factors in meson photoproduction needs to 
be revisited in the future.

\subsection{Other isospin channels}
\begin{figure}[t]
\centering\psfig{file=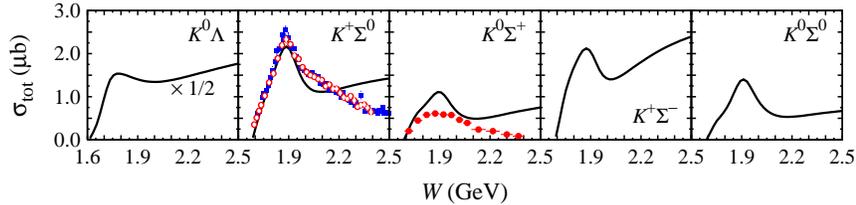,width=4.5in}
\caption{Total cross sections for the rest 5 isospin channels
of kaon photoproduction on the nucleon. Solid lines show the
calculated cross sections of Kaon-Maid\,\cite{maid}, whereas
solid squares, open circles, and solid circles display experimental
data from Refs.\,\cite{bradford06}, \cite{glander04}, and 
\cite{lawall05}, respectively. Note that experimental data in 
the $K^+\Sigma^-$ channel are only available in the differential
cross section\,\cite{kohri06}, whereas the cross section of the
$K^0\Lambda$ channel has been renormalized by a factor of 1/2
in order to fit on the scale.}
\label{fig:iso}
\end{figure}

By considering the conservation of quantum numbers, there 
are six possible reaction channels for kaon photoproduction on the 
nucleon, three on the proton and three on the neutron. 
Except for the $\gamma p\to K^+\Lambda$ channel, these possible
reactions are shown in Fig.~\ref{fig:iso}, where the calculations
obtained from Kaon-Maid are compared with the available data.
Note that there are few data points available just recently for
the $\gamma n\to K^+\Sigma^-$ differential cross section released by the
LEPS collaboration\,\cite{kohri06}. Thus, there are no data
for the neutral kaon photoproduction on the neutron,
$\gamma n\to K^0\Lambda$ and $\gamma n\to K^0\Sigma^0$,
although from Fig.~\ref{fig:iso} it is clear that the
data can severely constrain the model, especially in the 
$\gamma n\to K^0\Lambda$ channel. Furthermore, the
investigation of the $YN$ potentials through the
$\gamma d\to KYN$ reaction requires the information from these two
channels\,\cite{salam06}. We note that important progress in this direction 
has been indicated by the newly published  $\gamma d\to K^0KY$ data
by the Tohoku's group\,\cite{Tsu08}. 
It is also important to note that the $K^0$ photoproduction
excludes the $t$-channel in the Born terms. Thus, by comparing 
photoproductions of $K^+$ and $K^0$, one can study the influence
of the kaon propagator in kaon photoproduction.

\subsection{Kaon electroproduction}
\begin{figure}[t]
\centering\psfig{file=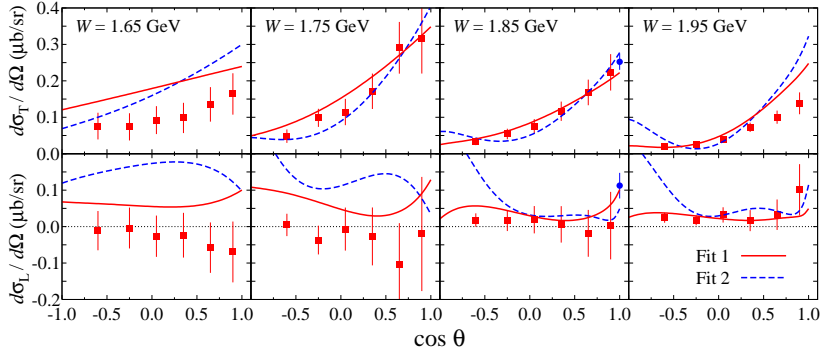,width=4.3in}
\caption{Longitudinal and transverse differential cross sections
  of the $ep\to e'K^+\Lambda$ process. The CLAS data 
  (solid squares\,\protect\cite{ambrozewicz} and 
  solid circles\,\protect\cite{mohring03}) are compared with
  the results from two multipoles models Fit 1 and Fit 2 (see
  Ref.~\protect\cite{mart_hyp06} for details).}
\label{fig:electro}
\end{figure}

It is widely known that the electroproduction process can give 
more information not available from the photoproduction one. 
One of the important
information  is the electromagnetic structure 
(form factor) of the kaon, which is hiding in the
longitudinal cross section ($d\sigma_{\rm L}/d\Omega$).
In contrast to the pion case, where the mass of pion is much
smaller and therefore the $t$-channel can dominate the process,
so that an independent extraction is possible\,\cite{Mart:2008sw},
the extraction of kaon electromagnetic form factor requires 
a reliable phenomenological model. Consequently, accurate
measurements of $d\sigma_{\rm L}/d\Omega$ are
required for this purpose. However, as seen from the 
lower panels of Fig.~\ref{fig:electro}, the presently 
available data still do not allow for the extraction of this 
form factor. Furthermore, to suppress contaminations
from nucleon resonances, measurements with $W\ge 2.4$ GeV 
are recommended.

\section{Summary and Conclusion}
The recent progress and some important issues in 
photo- and electroproduction of kaon on the nucleon have been
briefly presented. It is quite apparent that more experimental
and theoretical works are needed to settle the present problems
in this field. Nevertheless, with the modern concept of accelerator 
and spectrometer technologies available at JLAB, MAMI, ELSA, 
SPRING8, and other laboratories, we are quite optimistic that
these problems can be solved in the near future.

\section*{Acknowledgments}
The author acknowledges the support from the 
Universitas Indonesia.

\end{document}